# A direct basis approach to nonorthogonality in second quantization. Theory and application

Zixuan Hu, Mark A. Ratner, and Tamar Seideman[1]

*Department of Chemistry, Northwestern University, Evanston, Illinois 60208-3113*

We present a direct basis formalism for using nonorthogonal basis sets in the second quantization framework. As an alternative to the dual basis formalism, a direct basis retains the Hermiticity relation between the creation and annihilation operators, with which the form of quantum operators -- e.g. the number operator and the Hamiltonian -- can be readily interpreted and manipulated. To tackle the difficulty of formulating quantum operators in the direct basis, we introduce the coefficient matrix and the generalized creation and annihilation operators, with which an arbitrary N-particle operator can be generated by simple matrix manipulations with the metric tensor of a general basis set. We illustrate the application of the direct basis formalism with the Hubbard Hamiltonian and a dynamical study with the Heisenberg equations of motion.

## I. Introduction

The use of the second quantization framework is ubiquitous in numerous areas that involve many-body quantum systems. In the majority of the studies the creation and annihilation operators are defined for an orthogonal basis set in the Fock space, which guarantees the Hermiticity relation and the simple anti-commutation relation for fermions (or commutation relation for bosons; to simplify the notation, we will only discuss fermions in the following, but the results can be easily extended to the case of bosons):

$$c_i = \left( c_i^\dagger \right)^\dagger$$
$$\left\{ c_i^\dagger, c_j \right\} = \delta_{ij} \tag{1}$$

where $c_i^\dagger$ and $c_i$ are the creation and annihilation operators of the basis vector $\phi_i$. In many situations however, the use of nonorthogonal basis sets is desired. In molecular and solid state physics, the localized atomic orbitals are often used as basis functions to represent the system and the Hamiltonian. In electron transport theory, partitioning of space into the electrodes and scattering regions requires the use of localized functions that are nonorthogonal[1]. To adapt the second quantization formalism to nonorthogonal basis sets, it is necessary to relax one of the conditions in Eq. (1)[2] -- i.e. either the Hermiticity or the anti-commutation relation needs to be changed. Due to algebraic difficulties caused by changing the anti-commutation relation, previous studies have always used the dual basis formalism, which removes the Hermiticity relation[1, 3-8]. The price of removing the Hermiticity relation between the creation and annihilation operators is that the number operator is no longer Hermitian, and hence cannot be interpreted directly. In addition, in the dual basis representation quantum operators such as the Hamiltonian are partitioned into non-Hermitian components, and cannot be readily tailored for approximations. On the other hand, the direct basis formulation retains the Hermiticity, but changes the anti-commutation relation to:

$$\left\{ a_i^\dagger, a_j \right\} = S_{ji} = \left\langle \phi_j \middle| \phi_i \right\rangle \tag{2}$$

---

[1] Author to whom correspondence may be addressed. E-mail Address t-seideman@northwestern.edu

The loss of the simple anti-commutation relation causes complexity in the evaluation of quantum operators, but the benefit of retaining the Hermiticity relation is obvious: quantum operators are naturally partitioned into Hermitian parts that can be readily augmented or truncated for approximation purposes. In addition, the results for particle populations are also directly interpretable due to Hermiticity of the number operator, which makes the direct basis formalism suitable for dynamical studies on particle transfers[9, 10].

In the following sections we present a direct basis formalism for using nonorthogonal basis sets in the second quantization framework. First, the creation and annihilation operators are explicitly defined for the direct basis by their actions on a Slater determinant. Second, the physical interpretation of the number operator within a general basis is considered. Next the coefficient matrix and the generalized creation and annihilation operators are introduced, with which a simple form for an arbitrary N-particle quantum operator can be established with the aid of the metric tensor **g** of a general basis. Finally an application of the direct basis formalism is illustrated with the formulation of the Hubbard Hamiltonian in a general basis and the propagation of electron dynamics with the Heisenberg equations of motion.

## II. Theory

### A. Number operator in a general direct basis.

To define a second quantization formalism for fermions, the creation operator $C_m$ for a basis vector $\phi_m$, the annihilation operator $A_n$ for a basis vector $\phi_n$, and the anticommutation relation $\left\{ A_n, C_m \right\}$ need to be explicitly defined. The conventional formalism assumes an orthogonal basis set, and has $C_m = A_m^\dagger$ and $\left\{ A_n, C_m \right\} = \delta_{mn}$, allowing us to write the creation operator as $a^\dagger$, the annihilation operator as $a$, and $\left\{ a_n, a_m^\dagger \right\} = \delta_{mn}$. In a nonorthogonal basis, it can be shown that either the Hermitian relation $C_m = A_m^\dagger$ or the anticommutation relation $\left\{ A_n, C_m \right\} = \delta_{mn}$ has to be given up. The dual basis formalism proposed in previous studies[5] maintains $\left\{ A_n, C_m \right\} = \delta_{mn}$ and re-defines the Hermitian relation as $C^m = A_m^\dagger$, where $C^m$ is the creation operator of the dual basis $\phi^m$, and similarly $\left( A^m \right)^\dagger = C_m$. The advantage of using the dual basis is that the operator algebra is simplified since $\{A_n, C_m\} = \delta_{mn}$. However, the number operator in such case would be $A_m C_m$, whose Hermitian conjugate is $C_m^\dagger A_m^\dagger = A^m C^m$, which is the number operator of the dual basis $\phi^m$. Since $(A_m C_m)^\dagger \neq A_m C_m$, the number operator defined this way is no longer Hermitian and hence does not carry physical meaning. For many application in physics and chemistry the number operator is an essential quantity with physical interpretation that requires it to correspond to an observable Hence an important goal of the alternative formalism to be proposed below is to maintain $C_m = A_m^\dagger$ and the Hermiticity of the number operator. To define the operators explicitly, we consider a single Slater determinant $|\Psi\rangle = |\psi_1, ..., \psi_N\rangle$ and define $a^\dagger(\varphi)|\Psi\rangle = |\varphi, \psi_1, ..., \psi_N\rangle$, where $a^\dagger(\varphi)$ is the creation operator of state $\varphi$. Now we want $(a^\dagger(\varphi))^\dagger = a(\varphi)$ to be the annihilation operator of state $\alpha$, so if we take the inner product with an arbitrary state $|\phi_1, ..., \phi_{N-1}\rangle$

$$
\begin{aligned}
\langle \phi_1, ..., \phi_{N-1} | a(\varphi) | \psi_1, ..., \psi_N \rangle &= \langle \psi_1, ..., \psi_N | a^\dagger(\varphi) | \phi_1, ..., \phi_{N-1} \rangle^* \\
&= \begin{vmatrix} \langle \psi_1 | \varphi \rangle & \langle \psi_1 | \phi_1 \rangle & ... & \langle \psi_1 | \phi_{N-1} \rangle \\ ... & ... & & ... \\ \langle \psi_N | \varphi \rangle & \langle \psi_N | \phi_1 \rangle & ... & \langle \psi_N | \phi_{N-1} \rangle \end{vmatrix}^* \\
&= \sum_{k=1}^{N} (-1)^{k-1} \langle \varphi | \psi_k \rangle \langle \phi_1, ..., \phi_{N-1} | \psi_1, ... \bar{\psi}_k ..., \psi_N \rangle
\end{aligned}
\tag{3}
$$

*where $\bar{\psi}_k$ means $\psi_k$ is missing from the determinant.* Observing (3) gives

$$
a(\varphi) | \psi_1, ..., \psi_N \rangle = \sum_{k=1}^{N} (-1)^{k-1} \langle \varphi | \psi_k \rangle | \psi_1, ... \bar{\psi}_k ..., \psi_N \rangle
\tag{4}
$$

Eq.(4) gives the explicit definition of $a(\varphi)$. It can be verified that $\{a_i^\dagger, a_j^\dagger\} = 0$, $\{a_i, a_j\} = 0$, and $\{a_i^\dagger, a_j\} = \langle \varphi_j | \varphi_i \rangle$. Now we will define the number operator $n(\varphi) = a^\dagger(\varphi) a(\varphi)$, which is obviously Hermitian. The physical meaning of $n(\varphi)$ requires a word here because it is not immediately obvious that it still represents the number of particles in the state $\varphi$. We first consider the effect of $n(\varphi)$ on a normalized Slater determinant $|\Psi\rangle = |\psi_1, ..., \psi_N\rangle$ in which $\{\psi_i\}$ is an orthonormal basis that does not necessarily contain $\varphi$:

$$\langle \Psi | n(\varphi) | \Psi \rangle = \langle \psi_1,...,\psi_N | a^\dagger(\varphi) \sum_{k=1}^{N} (-1)^{k-1} \langle \varphi | \psi_k \rangle | \psi_1,...\bar{\psi}_k...,\psi_N \rangle$$

$$= \langle \psi_1,...,\psi_N | \sum_{k=1}^{N} (-1)^{k-1} \langle \varphi | \psi_k \rangle | \varphi,\psi_1,...\bar{\psi}_k...,\psi_N \rangle \qquad (5)$$

$$= \sum_{k=1}^{N} \left| \langle \varphi | \psi_k \rangle \right|^2$$

From (5) we see that the expectation value of $n(\varphi)$ is the total probability of projection of $|\Psi\rangle = |\psi_1,...,\psi_N\rangle$ into the state $\varphi$, which is a natural generalization from the usual meaning of the number operator under an orthonormal basis that contains $\varphi$. Now suppose we have a more general normalized Slater determinant $|\Phi\rangle = A_{Norm} |\phi_1,...,\phi_M\rangle$ in which $\{\phi_i\}$ is a normalized but nonorthogonal basis whose span is the same as that of $\{\psi_i\}$,. A Slater determinant formed by a nonorthogonal basis may not itself be normalized even when the basis vectors are all normalized, thus a global normalization constant $A_{Norm}$ has been introduced. If $M \leq N$ we can always re-write $|\Phi\rangle$ in the orthonormal basis $\{\psi_i\}$ via $|\Psi_j\rangle = |\psi_{j1},...,\psi_{jM}\rangle$:

$$|\Phi\rangle = A_{Norm} |\phi_1,...,\phi_M\rangle$$

$$= \sum_{j=1}^{C(N,M)} A_j |\Psi_j\rangle \qquad (6)$$

where $C(N,M)$ is the number of combinations for selecting M basis vectors from $\{\psi_i\}$ (which has N basis vectors), $A_{Norm}$ has been absorbed into $A_j$, and $|\Psi_j\rangle = |\psi_{j1},...,\psi_{jM}\rangle$ is one particular Slater determinant composed of M basis vectors from the collection $\{\psi_i\}$, for example (with $|\Psi_j\rangle$ randomly ordered):

$$|\Psi_1\rangle = |\psi_1,...,\psi_M\rangle$$

$$|\Psi_2\rangle = |\psi_1,\bar{\psi}_2,\bar{\psi}_3,\psi_4...,\psi_{M+2}\rangle$$

$$|\Psi_3\rangle = |\psi_2,\psi_3,\bar{\psi}_4,\psi_5...,\psi_{M+2}\rangle \qquad (7)$$

$$... \text{ etc.}$$

where $\bar{\psi}_2,\bar{\psi}_3$ in $|\Psi_2\rangle$ mean that $\psi_2$ and $\psi_3$ are missing from the determinant and the same for $\bar{\psi}_4$ in $|\Psi_3\rangle$. Now since each $|\Psi_j\rangle$ is in the orthonormal basis $\{\psi_i\}$, we can use the result of (5) and obtain:

$$\begin{aligned}
\left\langle \Phi \middle| n(\varphi) \middle| \Phi \right\rangle &= \sum_{j=1}^{C(N,M)} \left| A_j \right|^2 \left\langle \Psi_j \middle| n(\varphi) \middle| \Psi_j \right\rangle \\
&= \sum_{j=1}^{C(N,M)} \left| A_j \right|^2 \sum_{k=1}^{M} \left| \left\langle \varphi \middle| \psi_{jk} \right\rangle \right|^2
\end{aligned} \tag{8}$$

Eq. (8) has a natural interpretation of the probability of projecting $\Phi$ onto $\varphi$ and we have established the physical interpretation of the number operator $n(\varphi)$ for an arbitrary basis.

*B. N-particle operator in a general direct basis.*

To develop an expression for an N-particle operator in second quantization with a general direct basis, we start from an arbitrary single-particle quantum operator $\hat{A}$. Under an orthonormal basis $\{\psi_i\}$, $\hat{A}$ can be expressed in the matrix form $\mathbf{A} = (A_{ij})$ where $A_{ij} = \left\langle \psi_i \middle| \hat{A} \psi_j \right\rangle$. In second quantization form, $\hat{A}$ can be expressed as $\hat{A} = \sum_{i,j} A_{ij} a_i^\dagger a_j$, where $a_i^\dagger$ and $a_j$ are the creation and annihilation operators respectively defined for $\psi_i$ and $\psi_j$. We note that in the second quantization form of $\hat{A}$, the coefficient before each component operator $a_i^\dagger a_j$ coincides with the matrix element $A_{ij} = \left\langle \psi_i \middle| \psi_j \right\rangle$ for the orthonormal basis $\{\psi_i\}$. For a non-orthogonal basis ,$\{\phi_i\}$ however, the coefficients need to be re-evaluated. Here we would like to use the direct basis to formulate an arbitrary quantum operator $\hat{A}$ with the aid of the metric tensor $\mathbf{g} = (S_{ij})$, where $S_{ij} = \left\langle \phi_i \middle| \phi_j \right\rangle$.

*We define a coefficient matrix $\bar{\mathbf{A}} = (\bar{A}_{mk})$, such that $\hat{A} = \sum_{m,k} \bar{A}_{mk} a_m^\dagger a_k$, where $a_m^\dagger$ and $a_k$ are the creation and annihilation operators respectively defined for $\phi_m$ and $\phi_k$ taken from a non-orthogonal basis $\{\phi_i\}$.*

To find the relation between $\bar{\mathbf{A}} = (\bar{A}_{mk})$ and $\mathbf{A} = (A_{ij})$, we explicitly evaluate $A_{ij} = \left\langle \phi_i \middle| \hat{A} \phi_j \right\rangle$:

$$\begin{aligned}
A_{ij} = \left\langle \phi_i \middle| \hat{A} \phi_j \right\rangle &= \sum_{mk} \left\langle \phi_i \middle| \bar{A}_{mk} a_m^\dagger a_k \phi_j \right\rangle \\
&= \sum_{mk} \left\langle a_m \phi_i \middle| \bar{A}_{mk} a_k \phi_j \right\rangle \\
&= \sum_{mk} S_{im} \bar{A}_{mk} S_{kj} \\
&= \left( \mathbf{g} \bar{\mathbf{A}} \mathbf{g} \right)_{ij}
\end{aligned} \tag{9}$$

where the second and third lines are evaluated easily by the explicit definition of the creation and annihilation operators in the direct basis. The last line of (9) is a very simple form that allows us to directly equate the matrices themselves, i.e.

$$\mathbf{A} = \mathbf{g}\bar{\mathbf{A}}\mathbf{g} . \qquad (10)$$

Here we have successfully related the coefficient matrix $\bar{\mathbf{A}} = \left(\bar{A}_{mk}\right)$ to the usual matrix form $\mathbf{A} = \left(A_{ij}\right)$. The advantage of the form in matrix multiplication is that all the index information is packaged in the matrix rules and not explicitly expressed, such that the simple form retains when we go to multi-particle operators where the maze of indices can often become daunting to navigate. The evaluation in (9) has explicitly used the single-particle assumption and does not directly apply to a multi-particle operator. We note however that the definition of the coefficient matrix $\bar{\mathbf{A}} = \left(\bar{A}_{mk}\right)$ is based on the natural ordering of the creation and annihilation operators of the single-particle basis, so it would be reasonable to conjecture that if we can define the creation and annihilation operators for the multi-particle basis, i.e. $b_i^\dagger$ would create the i$^{th}$ multi-particle basis vector and $b_j$ would annihilate the j$^{th}$ multi-particle basis vector (the terms "create" and "annihilate" should be interpreted loosely for the moment until explicit definitions are given in the following), then the result of (10) can be extended to any arbitrary multi-particle operators. In the following we proceed to prove this proposition as more formally stated:

For an arbitrary N-particle operator $\hat{B}$, if we define the generalized creation operator $b_l^\dagger$ and generalized annihilation operator $b_l$ for the N-particle basis $\{\Phi_l\}$ such that $b_l^\dagger = \frac{1}{\sqrt{N!}} a_{l_1}^\dagger a_{l_2}^\dagger ... a_{l_N}^\dagger$ and $b_l = \frac{1}{\sqrt{N!}} a_{l_N} a_{l_{N-1}} ... a_{l_1}$, where $a_{l_1}^\dagger$ through $a_{l_N}^\dagger$ and $a_{l_1}$ through $a_{l_N}$ are the creation and annihilation operators of the 1-particle basis vectors, then we can express the operator $\hat{B}$ as $\hat{B} = \sum_{k,l} \bar{B}_{kl} b_k^\dagger b_l$. Furthermore, if we define the coefficient matrix $\bar{\mathbf{B}} = \left(\bar{B}_{kl}\right)$, the operator overlap matrix $\mathbf{B} = \left(B_{ij}\right) = \left(\left\langle \Phi_i \middle| \hat{B}\Phi_j \right\rangle\right)$, and the N-particle basis metric tensor $\mathbf{G} = \left(T_{ij}\right) = \left(\left\langle \Phi_i \middle| \Phi_j \right\rangle\right)$, then $\mathbf{B} = \mathbf{G}\bar{\mathbf{B}}\mathbf{G}$.

To illustrate this result, we start from the simplest case scenario of a 2-particle operator $\hat{B}$. A 2-particle basis $\{\Phi_k\}$ can be written as the tensor product of 1-particle bases such as $\phi_i \otimes \phi_j$. Suppose the size of the 1-particle basis set is M, then we can order $\Phi_k$ with $\phi_i \otimes \phi_j$ such that $k = j + (i-1)M$, e.g. $\Phi_1 ... \Phi_M$ will be $\phi_1 \otimes \phi_1 ... \phi_1 \otimes \phi_M$, $\Phi_{M+1} ... \Phi_{2M}$ will be $\phi_2 \otimes \phi_1 ... \phi_2 \otimes \phi_M$, ..., $\Phi_{M^2-M} ... \Phi_{M^2}$ will be $\phi_M \otimes \phi_1 ... \phi_M \otimes \phi_M$. With this ordering of the basis, we obtain the relation between the 2-particle metric tensor $\mathbf{G} = T_{kl} = \left( \left\langle \Phi_k \middle| \Phi_l \right\rangle \right)$ and the 1-particle metric tensor $\mathbf{g} = \left( S_{ij} \right) = \left\langle \phi_i \middle| \phi_j \right\rangle$:

$$\mathbf{G} = \mathbf{g} \otimes \mathbf{g} \tag{11}$$

where the tensor product between two matrices uses the standard definition of Kronecker product in linear algebra. By (11), the indices of matrix elements in $\mathbf{G} = \left( T_{kl} \right)$ can be related to those of $\mathbf{g} = \left( S_{ij} \right)$ as

$$\begin{aligned} T_{kl} = \left\langle \Phi_k \middle| \Phi_l \right\rangle &= \left\langle \phi_i \otimes \phi_j \middle| \phi_m \otimes \phi_n \right\rangle \\ &= S_{im} S_{jn} \end{aligned} \tag{12}$$

where $k = j + (i-1)M$ and $l = n + (m-1)M$.

*We define the generalized annihilation operator $b_k$ such that its action on any basis $\Phi_l$ is :*

$$b_k \Phi_l = T_{kl} = \left\langle \Phi_k \middle| \Phi_l \right\rangle = S_{im} S_{jn} \tag{13}$$

then for any operator $\hat{B} = \sum_{\alpha, \beta} \bar{B}_{\alpha\beta} b_\alpha^\dagger b_\beta$, where $b_\alpha^\dagger$ is defined as the Hermitian conjugate of $b_\alpha$, we have:

$$\begin{aligned} B_{kl} = \left\langle \Phi_k \middle| \hat{B} \Phi_l \right\rangle &= \sum_{\alpha, \beta} \bar{B}_{\alpha\beta} \left\langle \Phi_k \middle| b_\alpha^\dagger b_\beta \Phi_l \right\rangle \\ &= \sum_{\alpha, \beta} \bar{B}_{\alpha\beta} \left\langle b_\alpha \Phi_k \middle| b_\beta \Phi_l \right\rangle \\ &= \sum_{\alpha, \beta} T_{k\alpha} \bar{B}_{\alpha\beta} T_{\beta l} \\ &= \left( \mathbf{G} \bar{\mathbf{B}} \mathbf{G} \right)_{kl} \end{aligned} \tag{14}$$

Hence we get the same matrix expression as in (10), i.e.

$$\mathbf{B} = \mathbf{G} \bar{\mathbf{B}} \mathbf{G} = (\mathbf{g} \otimes \mathbf{g}) \bar{\mathbf{B}} (\mathbf{g} \otimes \mathbf{g}) \tag{15}$$

To relate the operator $b_k$ to the 1-particle annihilation operators, we note that if $\Phi_l = \phi_m \otimes \phi_n$ and $\Phi_\lambda = \phi_n \otimes \phi_m$, then:

$$b_k \frac{1}{\sqrt{2}} \left( \Phi_l - \Phi_\lambda \right) = b_k \left| \phi_m, \phi_n \right\rangle_{fer}$$

$$= \frac{1}{\sqrt{2}} \left( S_{im} S_{jn} - S_{in} S_{jm} \right) \qquad (16)$$

$$= \frac{1}{\sqrt{2}} a_j a_i \left| \phi_m, \phi_n \right\rangle_{fer}$$

where $\left| \phi_m, \phi_n \right\rangle_{fer}$ denotes a fermionic state, i.e. a Slater determinant. We can see from (16) that $b_k = \frac{1}{\sqrt{2}} a_j a_i$ in the fermionic Fock space. Similarly:

$$b_k \frac{1}{\sqrt{2}} \left( \Phi_l + \Phi_\lambda \right) = b_k \left| \phi_m, \phi_n \right\rangle_{bos}$$

$$= \frac{1}{\sqrt{2}} \left( S_{im} S_{jn} + S_{in} S_{jm} \right) \qquad (17)$$

$$= \frac{1}{\sqrt{2}} a_j a_i \left| \phi_m, \phi_n \right\rangle_{bos}$$

where $\left| \phi_m, \phi_n \right\rangle_{bos}$ denotes a bosonic state, i.e. a Slater permanent. Again in (17) $b_k = \frac{1}{\sqrt{2}} a_j a_i$ in the bosonic Fock space. *Hence it is natural to define the action of $b_k$ to be $b_k = \frac{1}{\sqrt{2}} a_j a_i$ when restricted to a particular Fock space.* If we use the definition $b_k^\dagger = \left( b_k \right)^\dagger$, then $b_k^\dagger = \frac{1}{\sqrt{2}} a_i^\dagger a_j^\dagger$, with which the operator $\hat{B}$ becomes:

$$\hat{B} = \sum_{k,l} \bar{B}_{kl} b_k^\dagger b_l = \frac{1}{2} \sum_{ijmn} \bar{B}_{ijmn} a_i^\dagger a_j^\dagger a_n a_m \qquad (18)$$

$$k = j + (i-1)M, \quad l = n + (m-1)M$$

We see that Eq. (18) is similar to the conventional form of the 2-particle operator in terms of creation and annihilation operators of an orthonormal basis set, except that here the coefficient matrix $\bar{\mathbf{B}} = \left( \bar{B}_{kl} \right)$ has the relation in (15) to the operator overlap matrix $\mathbf{B} = \left( B_{ij} \right) = \left( \left\langle \Phi_i \middle| \hat{B} \Phi_j \right\rangle \right)$.

The advantage of defining the generalized creation and annihilation operators $b_k$ and $b_k^\dagger$ is that the proof above can be easily generalized to an N-particle operator. If we define $b_l^\dagger = \dfrac{1}{\sqrt{N!}} a_{l_1}^\dagger a_{l_2}^\dagger \ldots a_{l_N}^\dagger$ and $b_l = \dfrac{1}{\sqrt{N!}} a_{l_N} a_{l_{N-1}} \ldots a_{l_1}$ then we may write any N-particle operator as $\hat{B} = \sum_{k,l} \breve{B}_{kl} b_k^\dagger b_l$ and $\mathbf{\breve{B}} = \left( \breve{B}_{kl} \right)$ would have the similar relation as in (15):

$$\mathbf{B} = \mathbf{G}\mathbf{\breve{B}}\mathbf{G} = \mathbf{g}^{\otimes N} \mathbf{\breve{B}} \mathbf{g}^{\otimes N} \tag{19}$$

where $\mathbf{g}^{\otimes N}$ is the $N^{th}$ power in terms of the tensor product $\otimes$.

### C. Example: the Hubbard Hamiltonian

In this subsection we illustrate the application of the results above to the Hubbard Hamiltonian, which is widely used in the modeling of materials in many areas of physics:

$$\hat{H} = \left[ -t \sum_{i,\sigma} (a_{i,\sigma}^\dagger a_{i+1,\sigma} + h.c.) \right] + \left[ U \sum_i n_{i\uparrow} n_{i\downarrow} \right] \tag{20}$$

where $\sigma$ is the spin index, $n_{i\uparrow} = a_{i\uparrow}^\dagger a_{i\uparrow}$, $n_{i\downarrow} = a_{i,\downarrow}^\dagger a_{i,\downarrow}$, $U$ is the on-site Coulomb repulsion term, and $t$ is the hopping integral (we take the convention $t > 0$ such that there is a negative sign before it). The expression (20) assumes an orthonormal basis, and for a general basis we need to re-write the Hamiltonian in the general form:

$$\hat{H} = \left[ \sum_{i,j,\sigma} \bar{t}_{ij} (a_{i\sigma}^\dagger a_{j\sigma} + a_{j\sigma}^\dagger a_{i\sigma}) \right] + \left[ \sum_{ijmn} \breve{U}_{ijmn} a_{i\uparrow}^\dagger a_{j\downarrow}^\dagger a_{n\downarrow} a_{m\uparrow} \right]$$
$$= \hat{H}_{TB} + \hat{H}_C \tag{21}$$

Here the first part of (21), $\hat{H}_{TB}$, is the 1-particle tight-binding operator whose $N \times N$ coefficient matrix is $\mathbf{\breve{H}}_{TB} = \left( \bar{t}_{ij} \right)$; the second part, $\hat{H}_C$, is the 2-particle Coulomb operator whose coefficient tensor $\breve{U}_{ijmn}$ is the two-particle counterpart of the coefficient matrix $\mathbf{\breve{H}}_{TB} = \left( \bar{t}_{ij} \right)$ and can be converted into a $N^2 \times N^2$ coefficient matrix $\mathbf{\breve{H}}_C$ with proper indexing. $\breve{U}_{ijmn}$ will reduce to the on-site Coulomb repulsion $U$ when the basis is orthonormal and all the atomic sites have equal repulsions. Now let us first consider the tight-binding part. If the values of all entries of $\mathbf{H}_{TB} = \left( \left\langle \phi_i \middle| \hat{H}_{TB} \phi_j \right\rangle \right)$ are known, then direct application of (10) will give us:

$$\mathbf{H}_{TB} = \mathbf{g}\mathbf{\breve{H}}_{TB}\mathbf{g} \quad \Rightarrow \quad \mathbf{\breve{H}}_{TB} = \mathbf{g}^{-1}\mathbf{H}_{TB}\mathbf{g}^{-1} \tag{22}$$

In practice however, it is rare that the ab intio values of all entries of $\mathbf{H}_{TB} = \left( \left\langle \phi_i \middle| \hat{H}_{TB} \phi_j \right\rangle \right)$ are available. What is commonly used instead is an empirically determined Hamiltonian that gives energy levels that fit experimental data.

*We define the empirical Hamiltonian matrix $\tilde{\mathbf{H}}$ such that the solution to the eigenvalue equation $\tilde{\mathbf{H}}\psi = E\psi$ gives the energy levels that optimally fit experimental data.*

For the tight-binding part only, we note that the empirical $\tilde{\mathbf{H}}_{TB}$ is exactly the matrix form of the first (tight-binding) part in Eq. (20), because for an orthonormal basis the matrix form of the Schrodinger equation is an eigenvalue equation, i.e. $\tilde{\mathbf{H}}_{TB}\psi = E\psi$. Since we want to work with a general basis with the metric $\mathbf{g}$, we need a way to relate the empirical Hamiltonian matrix $\tilde{\mathbf{H}}_{TB}$ to the coefficient matrix $\bar{\mathbf{H}}_{TB}$. This is done by noticing that for the matrix $\mathbf{H}_{TB} = \left( \left\langle \phi_i \middle| \hat{H}_{TB}\phi_j \right\rangle \right)$, the Schrodinger equation has the form of a generalized eigenvalue equation:

$$\mathbf{H}_{TB}\psi = \mathbf{g}E\psi \tag{23}$$

For the same system Eq. (23) should give the same energy levels as $\tilde{\mathbf{H}}_{TB}\psi = E\psi$, hence immediately we have:

$$\begin{cases} \mathbf{H}_{TB} = \mathbf{g}\tilde{\mathbf{H}}_{TB} \\ \bar{\mathbf{H}}_{TB} = \mathbf{g}^{-1}\mathbf{H}_{TB}\mathbf{g}^{-1} = \tilde{\mathbf{H}}_{TB}\mathbf{g}^{-1} \end{cases} \tag{24}$$

Eq. (24) gives us the relation to determine the tight-binding part of the Hamiltonian in a general basis.

The determination of $\bar{\mathbf{H}}_C$ is entirely analogous to that of the tight-binding part. We either start with the ab initio $U_{ijmn} = \left\langle \phi_{i\uparrow}\phi_{j\downarrow} \middle| \hat{H}_C \phi_{m\uparrow}\phi_{n\downarrow} \right\rangle$, convert it to the matrix $\mathbf{H}_C$ with proper indexing and use

$$\bar{\mathbf{H}}_C = \left( \mathbf{g} \otimes \mathbf{g} \right)^{-1} \mathbf{H}_C \left( \mathbf{g} \otimes \mathbf{g} \right)^{-1} \tag{25}$$

or start with an empirical $\tilde{\mathbf{H}}_C$ such that $\tilde{\mathbf{H}}_C\psi = E\psi$ gives energy values that fit experimental data, in which case

$$\bar{\mathbf{H}}_C = \tilde{\mathbf{H}}_C \left( \mathbf{g} \otimes \mathbf{g} \right)^{-1} \tag{26}$$

## III. Application

A. Initial conditions.

To illustrate the application of the direct basis formalism, we consider a dynamical study of electron transport over a 2-site molecular chain using the Heisenberg equations of motion[9]. Here we use a simple case: a 2-dimensional Hilbert space with a general normalized basis $\{\phi_1, \phi_2\}$ with the overlap $\langle \phi_1 | \phi_2 \rangle = S \in [0,1)$ (For simplicity, here we have assumed $S$ is real). In practice $\phi_1$ and $\phi_2$ are usually the localized atomic orbitals. An orthonormal basis $\{\phi_+, \phi_-\}$ can be easily constructed from $\{\phi_1, \phi_2\}$ with $\phi_{\pm} = \dfrac{1}{\sqrt{2(1 \pm S)}}(\phi_1 \pm \phi_2)$. In practice $\phi_{\pm}$'s are usually molecular orbitals that are eigenfunctions of the Hamiltonian of interest. If we start with $\langle n_+ \rangle_0$ electrons occupying $\phi_+ = \dfrac{1}{\sqrt{2(1+S)}}(\phi_1 + \phi_2)$, and $\langle n_- \rangle_0$ electrons occupying $\phi_- = \dfrac{1}{\sqrt{2(1-S)}}(\phi_1 - \phi_2)$, the expectation values of the number operators in the states $\phi_1$ and $\phi_2$ are determined by expressing $a_{\pm}^{\dagger}$ and $a_{\pm}$ as $a_{\pm}^{\dagger} = \dfrac{1}{\sqrt{2(1 \pm S)}}(a_1^{\dagger} \pm a_2^{\dagger})$ and $a_{\pm} = \dfrac{1}{\sqrt{2(1 \pm S)}}(a_1 \pm a_2)$, and applying the transformation:

$$\begin{cases} a_1^{\dagger} = \dfrac{1}{\sqrt{2}}(\sqrt{1+S}\,a_+^{\dagger} + \sqrt{1-S}\,a_-^{\dagger}) \\ a_1 = \dfrac{1}{\sqrt{2}}(\sqrt{1+S}\,a_+ + \sqrt{1-S}\,a_-) \end{cases} \tag{27}$$

to find

$$a_1^{\dagger} a_1 = n_1 = \frac{1}{2}\Big((1+S)n_+ + (1-S)n_- + 2\sqrt{1-S}\,a_+^{\dagger}a_-\Big) \tag{28}$$

where $n_+ = a_+^{\dagger}a_+$ and $n_- = a_-^{\dagger}a_-$. Repeating the same evaluation to $a_2^{\dagger}$ and $a_2$, we find that $a_2^{\dagger}a_2 = n_2 = \frac{1}{2}\Big((1+S)n_+ + (1-S)n_- - 2\sqrt{1-S^2}\,a_+^{?}a_-\Big)$. To proceed we need to use physical intuition. Since the initial states $\phi_{\pm} = \dfrac{1}{\sqrt{2(1+S)}}(\phi_1 \pm \phi_2)$ both contain equal contributions from $\phi_1$ and $\phi_2$, we must have $\langle n_1 \rangle_0 = \langle n_2 \rangle_0$ as initial conditions, which gives us $\langle a_+^{\dagger}a_- \rangle_0 = 0$ and

$$\langle n_1 \rangle_0 = \langle n_2 \rangle_0 = \frac{1}{2}\Big((1+S)\langle n_+ \rangle_0 + (1-S)\langle n_- \rangle_0\Big) \tag{29}$$

If we have $\langle n_+ \rangle_0 = 1$ and $\langle n_- \rangle_0 = 0$ then $\langle n_1 \rangle_0 = \langle n_2 \rangle_0 = \dfrac{1+S}{2}$. Note that both $\langle n_1 \rangle_0$ and $\langle n_2 \rangle_0$ reduce to 0.5 when $S = 0$, which is the expected value if $\{\phi_1, \phi_2\}$ were orthonormal. For nonzero S, the orthogonal states value of 0.5 is corrected by the $\dfrac{S}{2}$ term that incorporates the overlap of $\phi_1$ and $\phi_2$. Note that $\langle n_1 \rangle_0 + \langle n_2 \rangle_0 = 1 + S > \langle n_+ \rangle_0 + \langle n_- \rangle_0$, which means that $\langle n_1 \rangle_0$ and $\langle n_2 \rangle_0$ cannot be simultaneously observed (unless $S = 0$), consistent with standard quantum mechanical laws. If we let the system evolve with the total number of electrons conserved, i.e. $\langle n_+ \rangle + \langle n_- \rangle = N$, we get:

$$\begin{aligned}
\langle n_1 \rangle + \langle n_2 \rangle &= \left( (1+S)\langle n_+ \rangle + (1-S)\langle n_- \rangle \right) \\
&= N + S\langle n_+ \rangle - NS + S\langle n_+ \rangle \\
&= 2S\langle n_+ \rangle + N - NS
\end{aligned} \tag{30}$$

Since $\langle n_+ \rangle \in [0,1]$,

$$\langle n_1 \rangle + \langle n_2 \rangle \in [N - NS, N + (2-N)S] \tag{31}$$

and hence when N=1, $\langle n_1 \rangle + \langle n_2 \rangle \in [1 - S, 1 + S]$, whereas when N=2, $\langle n_1 \rangle + \langle n_2 \rangle \in [2 - 2S, 2]$. We see that when N=1, the sum $\langle n_1 \rangle + \langle n_2 \rangle$ is not exactly 1, but bounded between 1-S and 1+S. But when N=2, the sum $\langle n_1 \rangle + \langle n_2 \rangle$ will not exceed 2, instead it is bounded below by 2-2S. The fact that $\langle n_1 \rangle + \langle n_2 \rangle$ does not exceed 2 is a manifestation of the Pauli exclusion principle (for the same reason N cannot be greater than 2). On the other hand, if we let S=0 (orthogonal basis), then $\langle n_1 \rangle + \langle n_2 \rangle = N$ always.

Now we can evaluate $a_1^\dagger a_2$ similarly and get:

$$a_1^\dagger a_2 = \frac{1}{2}\left( (1+S)n_+ - (1-S)n_- \right) \tag{32}$$

So if $\langle n_+ \rangle_0 = 1$ and $\langle n_- \rangle_0 = 0$ then $\langle a_1^\dagger a_2 \rangle_0 = \dfrac{1+S}{2}$, note that even for $S = 0$, $\langle a_1^\dagger a_2 \rangle_0 = \dfrac{1}{2} \neq 0$. The initial condition we have been using -- i.e. $\langle n_+ \rangle_0$ electrons occupying $\phi_+ = \dfrac{1}{\sqrt{2(1+S)}}(\phi_1 + \phi_2)$, and $\langle n_- \rangle_0$ electrons occupying $\phi_- = \frac{1}{\sqrt{2(1-S)}}(\phi_1 - \phi_2)$ -- start the dynamics with occupation on the linear combination of atomic orbitals, i.e. molecular orbitals $\phi_\pm$ that are eigenfunctions of the Hamiltonian. In the following we will refer to this as the MO initial condition. In practice there is another kind of initial condition that starts the dynamics with occupation on the atomic orbitals themselves, and we will refer to this as the AO initial condition. To illustrate this we consider the system prepared initially by distributing $N$ electrons such that $N_1$ are in $\phi_1$ and $N_2$ in $\phi_2$. Since $\langle \phi_1 | \phi_2 \rangle = S \in [0,1)$ we find for the initial values of the number operators

$$\begin{aligned} \langle n_1 \rangle_0 &= N_1 + N_2 |S|^2 \\ \langle n_2 \rangle_0 &= N_2 + N_1 |S|^2 \end{aligned} \qquad (33)$$

Eq.(33) can be verified by directly applying Eq. (5) to the mixture of $\phi_1$ and $\phi_2$. Note that if $N_1 = N_2 = \frac{N}{2}$ we have

$$\langle n_1 \rangle_0 = \langle n_2 \rangle_0 = \frac{N(1+|S|^2)}{2} \qquad (34)$$

which is notably different from Eq.(29), i.e. $\langle n_1 \rangle_0 = \langle n_2 \rangle_0 = \frac{1}{2}\big((1+S)\langle n_+ \rangle_0 + (1-S)\langle n_- \rangle_0\big)$. Indeed, in (29) the dependence on $S$ is linear, yet in (34) it is quadratic. We note that this is another effect of the nonorthogonality of the basis, which requires separate treatments for the MO and AO initial conditions. When $S \to 0$ both (29) and (34) converges to $\frac{N}{2}$, as expected. Since $\phi_1 = \frac{1}{\sqrt{2}}(\sqrt{1+S}\phi_+ + \sqrt{1-S}\phi_-)$ and $\phi_2 = \frac{1}{\sqrt{2}}(\sqrt{1+S}\phi_+ - \sqrt{1-S}\phi_-)$, we have $\frac{\langle n_+ \rangle}{\langle n_- \rangle} = \frac{1+S}{1-S}$. Also $\langle n_+ \rangle + \langle n_- \rangle = N$, so $\langle n_+ \rangle = \frac{(1+S)N}{2}$ and $\langle n_- \rangle = \frac{(1-S)N}{2}$. We thus obtain from (32) that $\langle a_1^\dagger a_2 \rangle = \frac{1}{2}\big((1+S)\langle n_+ \rangle - (1-S)\langle n_- \rangle\big) = SN$. Note that unlike the MO initial condition, for the AO initial condition $\langle a_1^\dagger a_2 \rangle$ is independent of the actual values of $\langle n_1 \rangle$ and $\langle n_2 \rangle$.

*B. Dynamic simulation with the Hubbard Hamiltonian of ethylene.*

In this section we present an application of the methods developed in Sec. II and III A, considering the electron dynamics of the ethylene molecule described by the empirical Hubbard Hamiltonian:

$$\tilde{H} = \left[ -t \sum_{i,\sigma} (a_{i,\sigma}^\dagger a_{i+1,\sigma} + h.c.) \right] + \left[ U \sum_i n_{i\uparrow} n_{i\downarrow} \right]$$

$$= \tilde{H}_{TB} + \tilde{H}_C \qquad (35)$$

where the parameters are for example taken to be $t = 0.12$ and $U = 0.48$ (atomic units are used throughout), and the creation and annihilation operators correspond to orthonormal basis states. The Hartree-Fock mean-field (HFMF) approximation allows us to write the Coulomb part as:

$$\tilde{H}_C = U \sum_i \left( n_{i\uparrow} \langle n_{i\downarrow} \rangle + n_{i\downarrow} \langle n_{i\uparrow} \rangle - \langle n_{i\uparrow} \rangle \langle n_{i\downarrow} \rangle \right) \qquad (36)$$

where $\langle n_{i\downarrow} \rangle$ and $\langle n_{i\uparrow} \rangle$ are the expectation values of the occupation numbers on the corresponding sites. If we further assume that the two spins are equivalent at all times, the Hamiltonian $\tilde{H}$ becomes a simple TB Hamiltonian for each single spin, but with variable chemical potentials:

$$\tilde{H} = -t \sum_i (a_i^\dagger a_{i+1} + h.c.) + U \sum_i \left( n_i \langle n_i \rangle \right) \qquad (37)$$

where $\langle n_i \rangle$ should be evaluated for each time step in the propagation of the Heisenberg equations of motion. In matrix form (37) becomes:

$$\tilde{\mathbf{H}} = \begin{pmatrix} U \langle n_1 \rangle & -t \\ -t & U \langle n_2 \rangle \end{pmatrix} \qquad (38)$$

Now we define the atomic $2p_z$ orbitals on the 2 atoms of the ethylene molecule to be $\phi_1$ and $\phi_2$. The overlap integral is taken to be $S = 0.3$ [11]. By Eq.(24) we have:

$$\bar{\mathbf{H}} = \tilde{\mathbf{H}} \mathbf{g}^{-1}$$

$$= \frac{1}{1 - S^2} \begin{pmatrix} U \langle n_1 \rangle & -t \\ -t & U \langle n_2 \rangle \end{pmatrix} \begin{pmatrix} 1 & -S \\ -S & 1 \end{pmatrix} \qquad (39)$$

$$= \frac{1}{1 - S^2} \begin{pmatrix} U \langle n_1 \rangle + St & -t - U \langle n_1 \rangle S \\ -t - U \langle n_2 \rangle S & U \langle n_2 \rangle + St \end{pmatrix}$$

Eq. (39) appears to be problematic because the off-diagonal entries of $\bar{\mathbf{H}} = \tilde{\mathbf{H}} \mathbf{g}^{-1}$ could be different if $\langle n_1 \rangle \neq \langle n_2 \rangle$, which implies that $\bar{\mathbf{H}}$ is not Hermitian. We note that in principle $\bar{\mathbf{H}}$ should be Hermitian because the Hamiltonian operator is Hermitian regardless of the basis representation used. The reason for the non-Hermiticity of $\bar{\mathbf{H}}$ is that the HFMF approximation has converted the on-site repulsion $U$ term into a variable chemical potential term, which is then incorporated into the TB Hamiltonian. To correct this problem we note that the HFMF approximation assumes that the variations of $\langle n_1 \rangle$ and $\langle n_2 \rangle$ from $\langle n_1 \rangle_0$ and $\langle n_2 \rangle_0$ are very small and hence we can write:

$$
\begin{aligned}
\bar{\mathbf{H}} &= \tilde{\mathbf{H}}\mathbf{g}^{-1} \\
&= \frac{1}{1-S^2}\left\{\begin{pmatrix} U\langle n_1 \rangle + St & -t - U\langle n_1 \rangle_0 S \\ -t - U\langle n_2 \rangle_0 S & U\langle n_2 \rangle + St \end{pmatrix} + \begin{pmatrix} 0 & -US\Delta n_1 \\ -US\Delta n_2 & 0 \end{pmatrix}\right\} \\
&\approx \frac{1}{1-S^2}\begin{pmatrix} U\langle n_1 \rangle + St & -t - U\langle n_1 \rangle_0 S \\ -t - U\langle n_2 \rangle_0 S & U\langle n_2 \rangle + St \end{pmatrix}
\end{aligned}
\tag{40}
$$

where $\bar{\mathbf{H}}$ is approximately Hermitian when $\langle n_1 \rangle_0 = \langle n_2 \rangle_0$. Indeed the empirical Hamiltonian $\tilde{H}$ assumes the initial condition that $\langle n_1 \rangle_0 = \langle n_2 \rangle_0$. We note that the anomaly in (39) and its approximate fix in (40) are entirely introduced by using an empirical Hamiltonian and the HFMF approximation, and hence is not a weakness of the theory presented in Sec. II.

An interesting point to consider is that if we start from the coefficient matrix $\bar{\mathbf{H}} = \begin{pmatrix} \alpha_1 & \beta \\ \beta & \alpha_2 \end{pmatrix}$, which is by definition Hermitian, and reverse the process in Eq. (39) by calculating $\tilde{\mathbf{H}}$ from $\bar{\mathbf{H}}$:

$$
\begin{aligned}
\tilde{\mathbf{H}} &= \bar{\mathbf{H}}\mathbf{g} \\
&= \begin{pmatrix} \alpha_1 & \beta \\ \beta & \alpha_2 \end{pmatrix}\begin{pmatrix} 1 & S \\ S & 1 \end{pmatrix} \\
&= \begin{pmatrix} \alpha_1 + \beta S & \alpha_1 S + \beta \\ \beta + \alpha_2 S & \alpha_2 + \beta S \end{pmatrix}
\end{aligned}
\tag{41}
$$

then $\tilde{\mathbf{H}}$ is not Hermitian when $\alpha_1 \neq \alpha_2$, but the eigenvalues of $\tilde{\mathbf{H}}$ must be real because $\tilde{\mathbf{H}}$ is the empirical Hamiltonian that fits experimental data. This result is clarified by a mathematical theorem that says: if $\mathbf{H}$ and $\mathbf{g}$ are Hermitian, and if $\mathbf{g}$ is positive definite, then $\tilde{\mathbf{H}} = \mathbf{g}^{-1}\mathbf{H}$ must have real eigenvalues[12]. We note that by Eq.(22) $\mathbf{H} = \mathbf{g}\bar{\mathbf{H}}\mathbf{g}$, so $\mathbf{H}$ is Hermitian if $\bar{\mathbf{H}}$ and $\mathbf{g}$ are both Hermitian, which is the case indeed. In addition, the metric tensor $\mathbf{g}$ must be positive definite for a Hilbert space, so $\tilde{\mathbf{H}} = \mathbf{g}^{-1}\mathbf{H}$ must have real eigenvalues. The possibility of $\tilde{\mathbf{H}}$ being non-Hermitian is interesting because it is unique to the nonorthogonal formalism. The non-Hermiticity of $\tilde{\mathbf{H}}$ is realized when $\alpha_1 \neq \alpha_2$, i.e. when we have unequal energy expectation values for the basis orbitals (atomic orbitals), in which case we expect to see a qualitative difference between the dynamics calculated by the nonorthogonal formalism and the conventional orthogonal model. In the case of ethylene described by a Hubbard Hamiltonian, if we use the HFMF approximation, then as in Eq. (39) we see that the inequality between $\alpha_1$ and $\alpha_2$ is caused by the difference between $U\langle n_1 \rangle$ and $U\langle n_2 \rangle$, hence we predict that an increasing $U$ should cause increasing difference between nonorthogonal and orthogonal treatments. This will be demonstrated in the dynamic simulations below.

First, we use the MO initial condition described in Sec. III A. Since the MO initial condition starts the system in a stationary state, there would be no electron population dynamics to observe unless an external disturbance is applied. In the following we apply a source and a drain similar to the setup in Ref. [9], where we add (subtract) a source $s$ (drain $d$) term in the Heisenberg equations of motion:

$$\frac{d}{dt} a_1^\dagger a_1 = \frac{i}{\hbar} \left[ H, a_1^\dagger a_1 \right] + s$$

$$\frac{d}{dt} a_2^\dagger a_2 = \frac{i}{\hbar} \left[ H, a_2^\dagger a_2 \right] - d$$

(42)

where if we write $\langle n_i \rangle = \langle a_i^\dagger a_i \rangle$, then $s = A\sin(\omega \cdot t) \cdot (1 + \langle n_1 \rangle_0 - \langle n_1 \rangle)$ and $d = A\sin(\omega \cdot t) \cdot (1 + \langle n_2 \rangle - \langle n_2 \rangle_0)$ when $A\sin(\omega \cdot t) > 0$; $s = A\sin(\omega \cdot t) \cdot (1 + \langle n_1 \rangle - \langle n_1 \rangle_0)$ and $d = A\sin(\omega \cdot t) \cdot (1 + \langle n_2 \rangle_0 - \langle n_2 \rangle)$ when $A\sin(\omega \cdot t) < 0$. This is because when $A\sin(\omega \cdot t) > 0$ the system receives (loses) electrons from (to) the external source (drain), hence increasing $\langle n_1 \rangle$ ($\langle n_2 \rangle$) would reduce (increase) the source (drain), and vice versa when $A\sin(\omega \cdot t) < 0$. In this work, the source (drain) is not calculated by the coupling to a plasmonic resonance, but rather set to be $A = 0.04$, $\omega = 0.157$ ( atomic units used throughout). The MO initial condition is set to be $\langle n_+ \rangle_0 = 1$ and $\langle n_- \rangle_0 = 0$ which translates to $\langle n_1 \rangle_0 = \langle n_2 \rangle_0 = \frac{1 + S}{2}$. We use the second quantized Hamiltonian defined by the coefficient matrix in Eq. (40) with $t = 0.12$, and calculate the dynamics with different $S$ and $U$ values. Figure 1 illustrates our results for the $U = 0$ case:

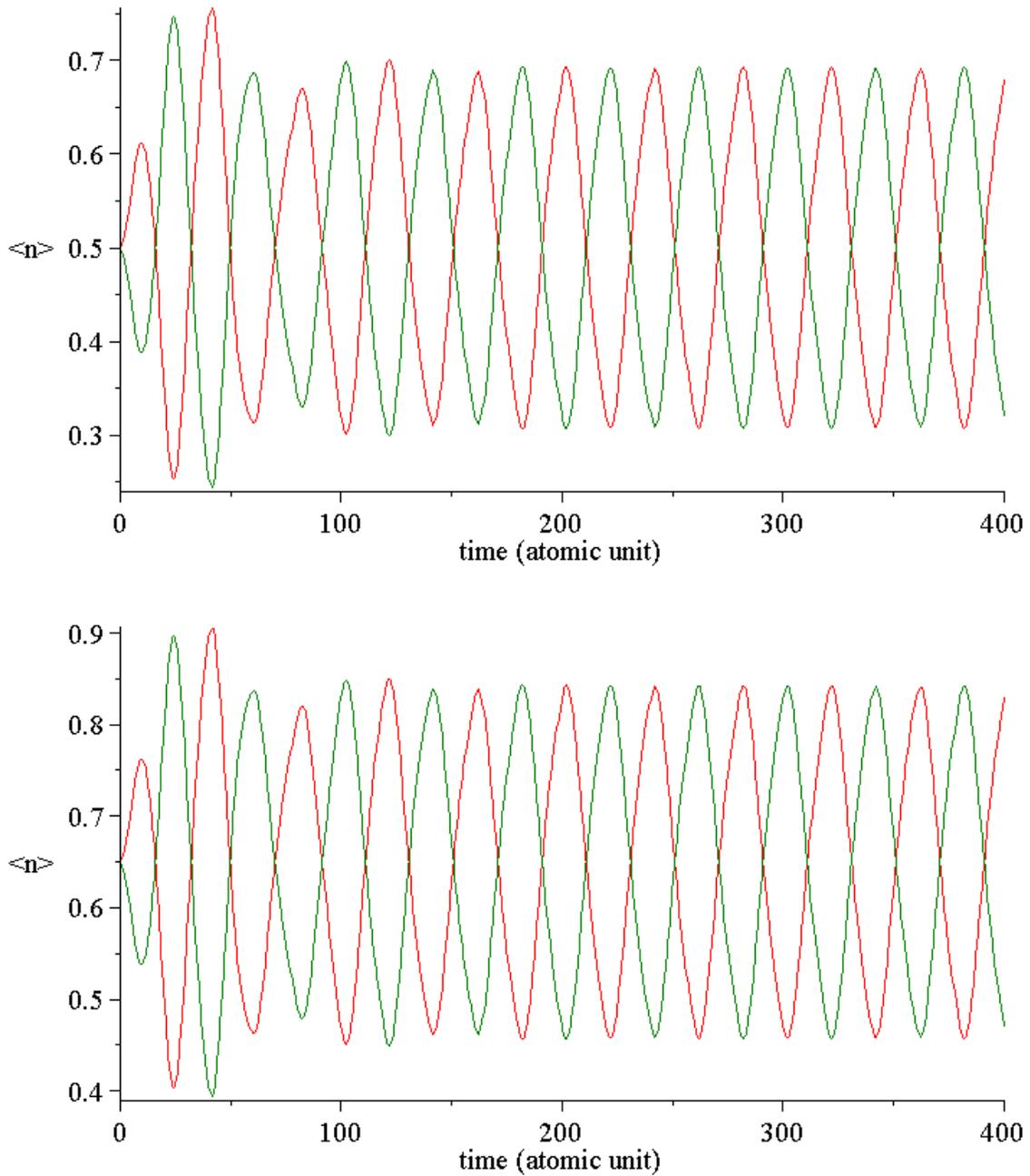

Figure 1. Electron population dynamics for ethylene with sinusoidal source and drain. Top figure is $U = 0$, $S = 0$; bottom is $U = 0$, $S = 0.3$. Red curve is $\langle n_1 \rangle$, green curve is $\langle n_2 \rangle$.

We see in Figure 1 that the dynamics calculated with $S = 0$ (top) is qualitatively similar to that calculated with $S = 0.3$ (bottom). The only difference is the values of $\langle n_1 \rangle$ and $\langle n_2 \rangle$, which are greater with $S = 0.3$ due to the interpretation of $\langle n \rangle$ -- the expectation value of the number operator -- defined in Eqs. (5) through (8).

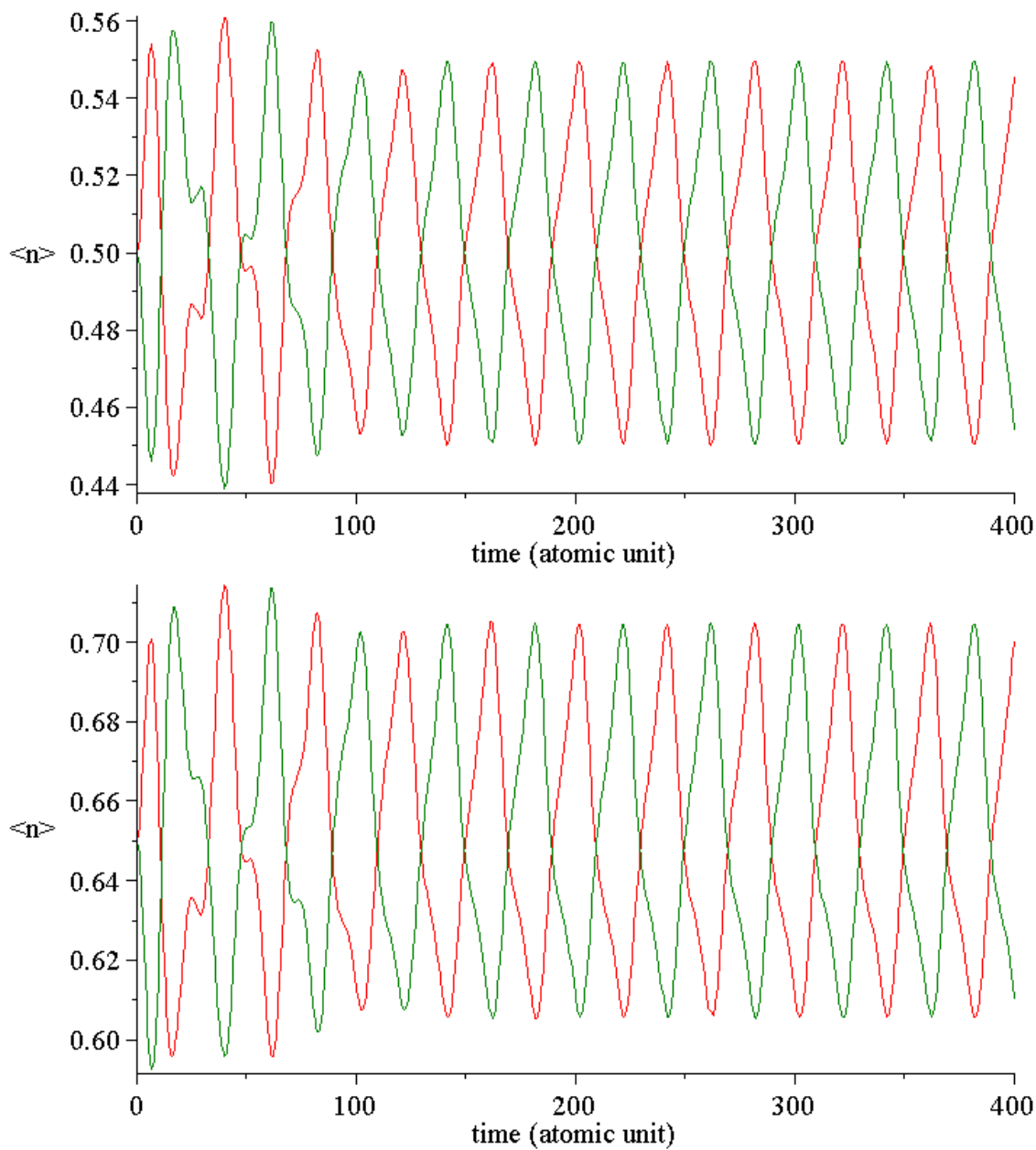

Figure 2. Electron population dynamics for ethylene with sinusoidal source and drain. Top figure is $U = 0.48 = 4t$, $S = 0$; bottom is $U = 0.48$, $S = 0.3$. Red curve is $\langle n_1 \rangle$, green curve is $\langle n_2 \rangle$.

In Figure 2 we increase $U$ to the empirical value 0.48 (4t) which results in two major effects. First is a significant reduction in the amplitude of the oscillations for $\langle n_1 \rangle$ and $\langle n_2 \rangle$, which is reasonably attributed to the on-site repulsion. Second is a qualitative profile difference between the curves calculated with $S = 0$ (top) and $S = 0.3$ (bottom): the oscillation in the top is symmetric, while the oscillation in the bottom is asymmetric.

We suggest that this difference in profile is caused by the non-Hermitian effect that we have predicted in Eq. (41) and in the arguments that immediately follow. Further increase of $U$, is expected to enhance the asymmetry in the oscillations and Figure 3 illustrates that this is the case,

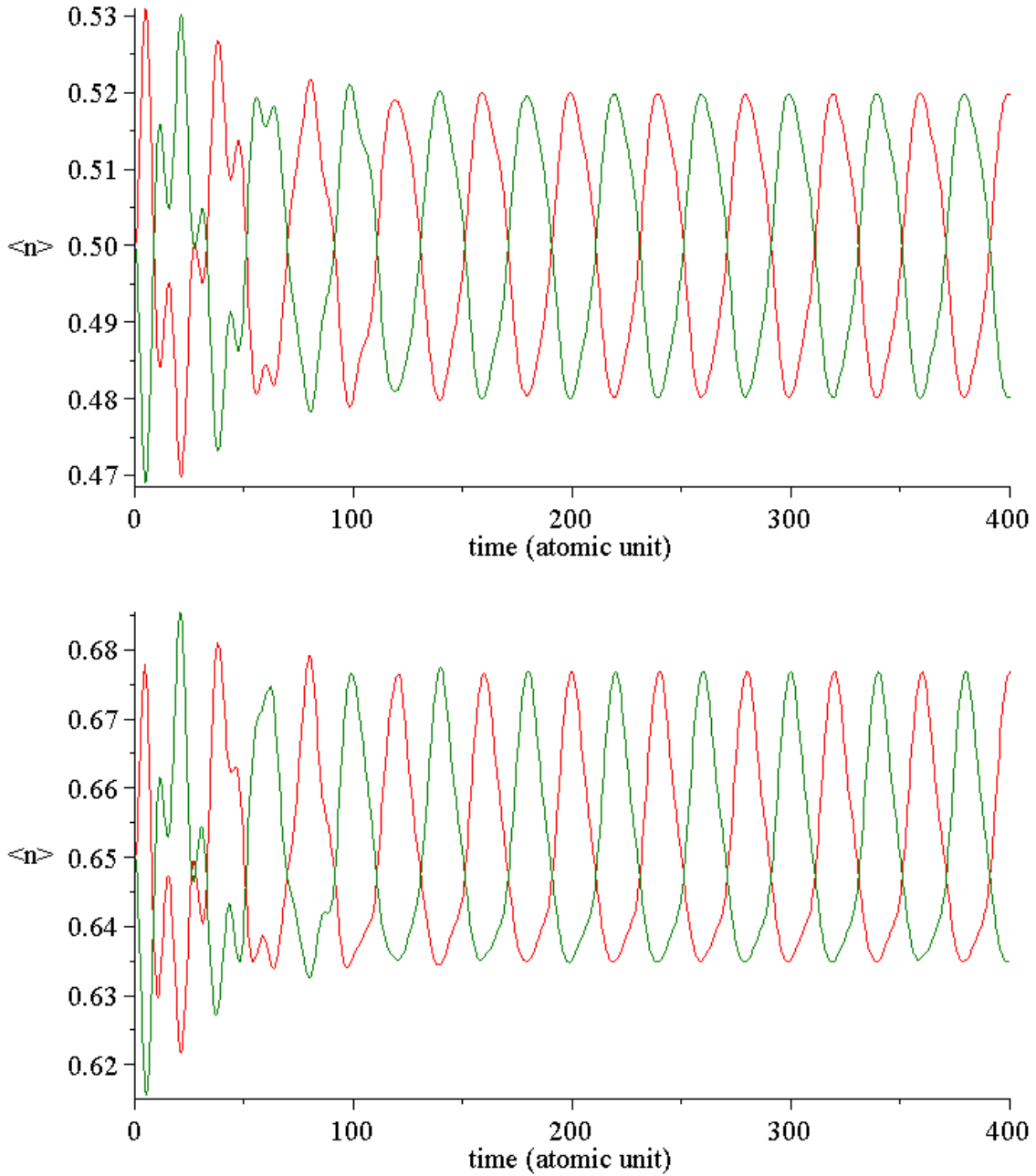

Figure 3. Electron population dynamics for ethylene with sinusoidal source and drain. Top figure is $U = 1.2 = 10t$, $S = 0$; bottom is $U = 1.2$, $S = 0.3$. Red curve is $\langle n_1 \rangle$, green curve is $\langle n_2 \rangle$.

Figures 1-3 together support our arguments below Eq. (41) that an asymmetry in the diagonal terms of the TB Hamiltonian will cause qualitative difference between the dynamics calculated with orthogonal and non-orthogonal models. Since in the current case of ethylene, the asymmetry is caused by the difference between $U\langle n_1 \rangle$ and $U\langle n_2 \rangle$, an increased $U$ has caused greater differences between the dynamic profiles with $S = 0$ and $S = 0.3$.

Finally we consider the AO initial condition described in Sec. III A. For a general AO condition the system starts in a non-stationary state, and the ensuing dynamics will be sinusoidal without any external disturbance. Since the variations of $\langle n_1 \rangle$ and $\langle n_2 \rangle$ are expected to be substantial, we cannot make the approximation in Eq. (40), and will only use the TB part of the Hamiltonian (obtained by setting $U = 0$ in Eq. (40)) in the following. We set the initial condition to be $N_1 = 1$ and $N_2 = 0$, which by Eq. (33) translates to $\langle n_1 \rangle_0 = 1$ and $\langle n_2 \rangle_0 = S^2$. For this simple case the time evolution of $\langle n_i \rangle$ can be calculated analytically to be

$$\langle n_1 \rangle = \frac{1 + S^2}{2} + \frac{1 - S^2}{2} \cos 2\beta t$$
$$\langle n_2 \rangle = \frac{1 + S^2}{2} - \frac{1 - S^2}{2} \cos 2\beta t \qquad (43)$$

where we have replaced the $t$ in the Hubbard Hamiltonian with $\beta$ to avoid confusion with the time variable. The analytic expressions show that S has quantitative effects on the average and the amplitude of the oscillations. The comparison is shown in Figure 4.

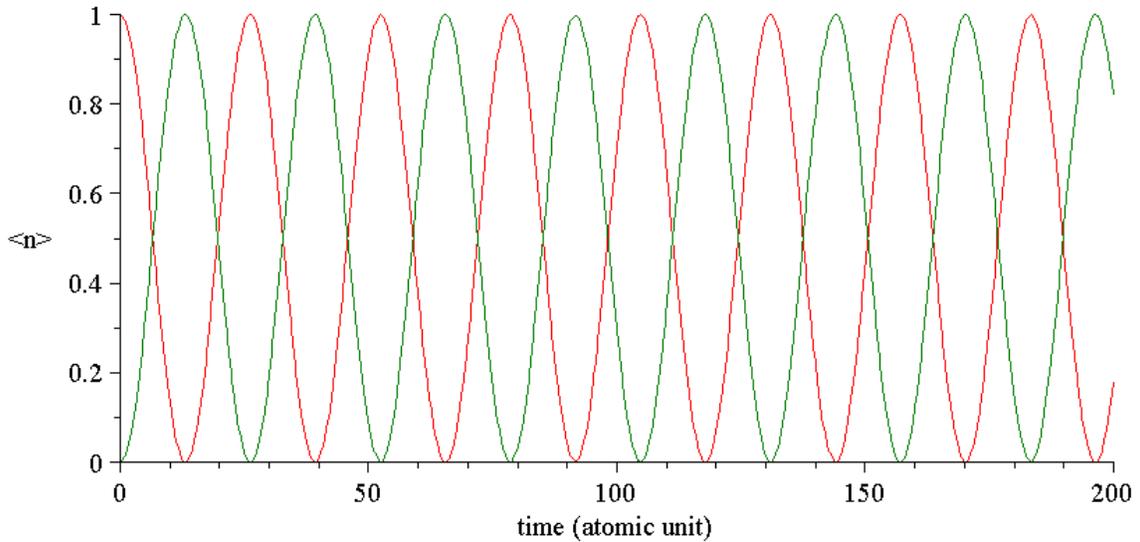

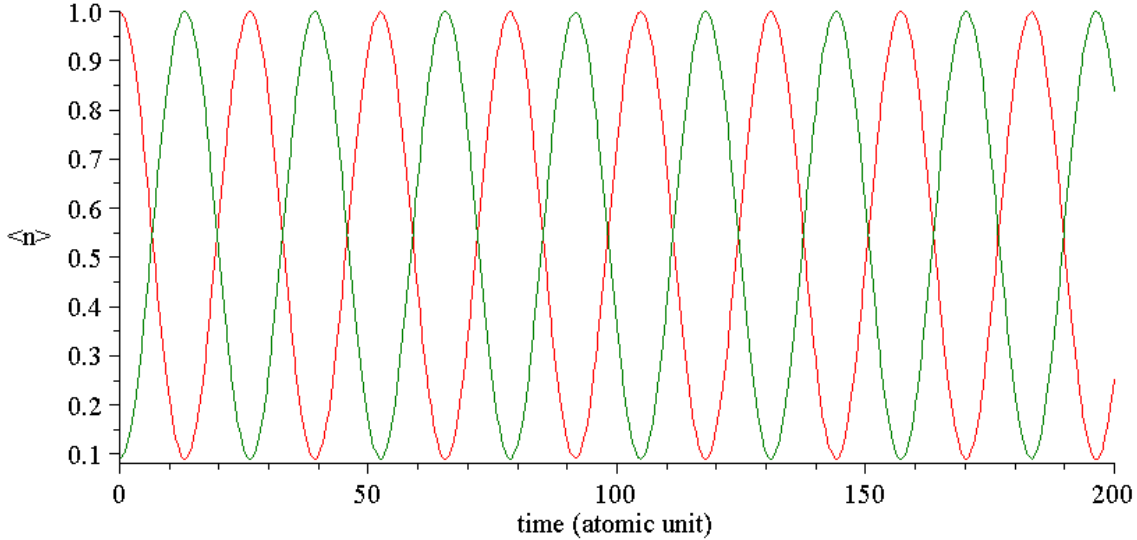

Figure 4. Electron population dynamics for ethylene with AO initial condition. Top figure is $U = 0$, $S = 0$; bottom is $U = 0$, $S = 0.3$. Red curve is $\langle n_1 \rangle$, green curve is $\langle n_2 \rangle$.

We see in Figure 4 that oscillations in the two panels share the same single frequency profile determined by $\cos 2\beta t$. The average and amplitude of the oscillations are as predicted by the analytic expressions. Since the correction caused by a finite $S$ scales as $S^2$ in (43), the correction is much less significant than that for the MO initial condition, where the correction scales as $S$. In Figure 4 there is no qualitative difference between the top and bottom figures because for simplicity we have set $U = 0$ and hence the non-Hermitian behavior predicted by (41) is not seen. We note that for an asymmetric molecule there may be inherent imbalance in the chemical potentials of different atomic sites, and hence qualitative difference may arise for a finite $S$ versus $S = 0$. We will investigate the dynamics of asymmetric molecules in a future report.

## IV. Conclusion

In this report we developed and applied a direct basis formalism for the second quantization formulation in a nonorthogonal basis. The direct basis formalism retains the Hermiticity relation between the creation and annihilation and hence offers physical interpretation of the number operator and tailoring of the Hamiltonian. With the aid of the metric tensor and the generalized creation and annihilation operators, the second quantized form of a general N-particle operator in a nonorthogonal basis was derived. Several concerns that arise from the application of the abstract theory were discussed, such as the different types of initial conditions, and the non-Hermiticity of the empirical Hamiltonian caused by an imbalance in the diagonal elements of the TB Hamiltonian. Finally a sample simulation has been carried out on the charge dynamics of ethylene described by a Hubbard Hamiltonian.